\documentclass[%
 reprint,
 amsmath,amssymb,
 aps,
]{revtex4-2}

\usepackage{dcolumn}
\usepackage{bm}

\usepackage{graphicx}
\usepackage{subfigure}
\usepackage{graphicx,hyperref}
\expandafter\let\csname equation*\endcsname\relax
\expandafter\let\csname endequation*\endcsname\relax
\usepackage{amsmath}
\usepackage{amssymb}
\usepackage{multirow, makecell, comment} 
\newcommand{\fla}[1]{\begin{flalign}#1\end{flalign}}
\usepackage{braket}
\usepackage{lineno}
\usepackage{lipsum}
\usepackage{xcolor}
\usepackage{soul}
\usepackage{array}

\begin{document}

\preprint{APS/123-QED}
\title{Area theorem for surface plasmons interacting with resonant atoms}

\author{Sergey A. Moiseev$^{1 }$, and Ali. A. Kamli$^{2 }$}

\affiliation{$^{1}$Kazan Quantum Center, Kazan National Research Technical University n.a. A.N. Tupolev-KAI, 10 K. Marx St., 420111, Kazan, Russia}
\affiliation{$^{2}$ Department of Physics, Jazan University, Jazan Box 114, Saudi Arabia.
\\
 E-mails: s.a.moiseev@kazanqc.org, alkamli@jazanu.edu.sa}

\date{\today}

\begin{abstract}
We show how the area theorem is applicable to the analytical description of the nonlinear interaction of surface plasmon modes with resonant two-level atoms. 
A closed analytical solution is obtained, which shows that surface plasmons can form long-propagating $2\pi$ pulses when interacting with an optically dense two-level atomic ensemble. The possible applications of the surface pulse area theorem and the conditions for the detection of $2\pi$ surface plasmon pulses are discussed.

\end{abstract}

\maketitle

\section{Introduction}

The resonant interaction of light pulses with coherent atomic ensembles plays an important role in quantum optics, laser physics and quantum technologies \cite{Allen75,Scully97,Boyd03,Sang11}.
Considerable advances have been achieved over the past years in cavity systems and photonic structures and more recently in the field of plasmonics.
The surface plasmon (SP) modes propagate at the interface connecting a dielectric and metal with highly confined electric field amplitude and thus large energy concentration at nano scale distances near the interface \cite{Barnes03,Zayats05,Maier07,Sarid10,Ozbay06,Gram10,Su18,Atia19,Said20} 
The nano optics nature of these SP modes enable them to circumvent the diffraction limit, that other techniques suffer, making plasmonics strong candidates to generate strong coupling with atoms and thus provides potential platform to explore coherent light-atom interactions and possible device applications that require strong coupling with atoms. 
These properties of SP fields are of great interest for the coherent  interaction of light with resonant atomic ensembles \cite{Kamli08,Moiseev10,Kamli2011,Siomau12,Tan14,Asgar18,Asgar2021,Gu20,Tan21,Tian22,Duan22}.
The highly inhomogeneous spatial structure of the electromagnetic field of SP greatly complicates the theoretical description of the nonlinear effects of their interaction with atoms. 
These difficulties are currently being overcome by the use of perturbation theory \cite{Tan14,Asgar18,Gu20}, which has a limited scope of applicability and does not allow a more complete understanding of the nonlinear patterns of the interaction of SP pulses with resonant atoms.
In this paper, we want to draw attention to the usefulness of using the area theorem for the theoretical description of the nonlinear coherent interaction of SP pulses with resonant atoms.

The area theorem provides researchers with a powerful tool for obtaining exact analytical solutions of nonlinear equations \cite{MacCall69} that allow the description of general patterns of interaction and propagation of light pulses in a coherent resonant medium.
The area theorem was derived in the well-known work of McCall and Hahn \cite{MacCall69} for the propagation of a light pulse through a resonant medium of two-level atoms, 
what led to prediction of self-induced transparency for light pulses having a pulse area equal to $2\pi$,
 which then led to the discovery of optical solitons \cite{Lamb71}.
The pioneering theoretical and experimental work on area theorem \cite{MacCall69} for the interaction of a single light pulse with resonant atoms,
has induced much work devoted to its application to solving various problems of the interaction of light pulses with resonant atomic ensembles.
As an example, we note only the application of the area theorem
\cite{MacCall69,Lamb71,Ablowitz1974,Eberly98} to the photon echo in an optically depth medium  \cite{Hahn71,Fr71,Moiseev87,Urman,Moiseev20,Wang2022}
and in resonators \cite{Thierry2014,Moiseev22}, the cavity assisted Dicke superradiance \cite{Gr01}
, to the description of the interaction of light pulses with three-level \cite{Eberly02,Shch15} and four-level atoms \cite{Gut16} 
optical quantum memory protocols \cite{Moiseev04,Moiseev23}, 
description of various modes of laser generation \cite{Arkh16,Arkh21,Pakhom2023}.

In this work, we develop pulse area approach for SP modes interacting with two-level atomic ensemble near negative index meta-material (NIMM) \cite{Vesel68,Pendry96,Shalaev07} due to their ability to support both transverse electric (TE) and transverse magnetic (TM) polarized SP modes with low loss and energy confined to the interface. The interest in these artificially fabricated materials has been enhanced more recently by the fabrication of NIMMs at optical and desirable frequencies, the emergence of quantum plasmonics, and the possibilities of implementing optical quantum processing (see reviews \cite{Uriri18,Li20,Uriri20,Rivera20,Solntsev2021,Liu21}).
These SP modes couple strongly to inhomogeneously broadened two-level atoms, which results in nonlinear dynamics for the equations of motion of the SP modes.
We are interested in SP pulse area theorem and in formation of $2\pi$ short SP pulses in their resonant interaction with two-level atoms and their ability to modify 
the McCall and Hahn area theorem \cite{MacCall69}. 
Finally we discuss the obtained results and  its potential applications.  

\section{Plasmonic modes}

 In Fig. 1 we show the studied hybrid structure, containing  a dielectric/metamaterial interface with ensemble of two-level atoms.
 The upper half space of the interface ($z>0$) can be any dielectric material characterized by constant dielectric function or permittivity $\varepsilon_1$  and constant magnetic permeability $\mu_1$.
 In the analysis below we shall take the upper half to be a rare-earth crystal. 
 The two-level atoms are embodied in the dielectric crystal part of this structure with a given atomic density $\rho(x,z)$ and inhomogeneous broadening function $G(\Delta/\Delta_{in})$ of resonant transition  ($\Delta_{in}$ - linewidth) and detuning $\Delta$.
The lower half space of the interface ($z<0$) is a negative index meta-material (NIMM) characterized by frequency dependent complex dielectric function or permittivity $\varepsilon_2(\omega)$  and complex magnetic permeability $\mu_2(\omega)$. 
For certain frequency range, the real parts of $\varepsilon_2(\omega)$ and $\mu_2(\omega)$  are negative. 
When both real parts of $\varepsilon_2(\omega)$ and $\mu_2(\omega)$  are negative, both transverse electric (TE) and transverse magnetic (TM) polarized plasmonic modes can exist with low losses, and this is the main advantage of using NIMM in this work.  
\begin{figure}
\includegraphics[width=0.8\linewidth]{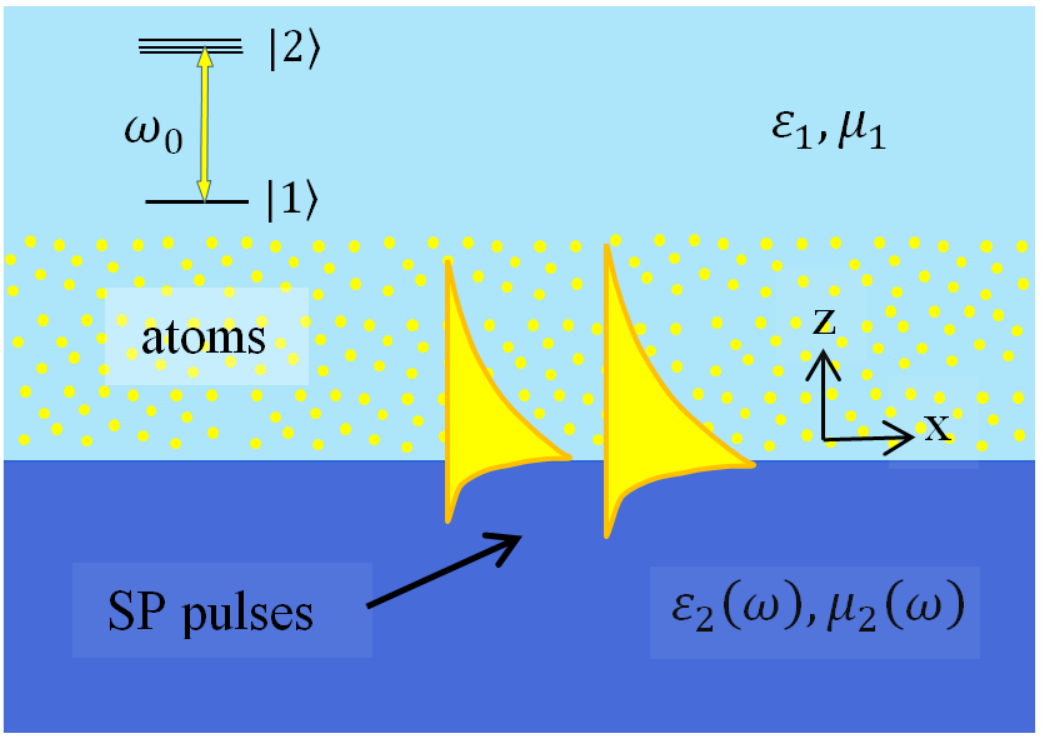}
\caption{Plasmonic structure  supporting SP modes consisting of upper half space ($z > 0$) of permittivity $\epsilon_1$ and permeability $\mu_1$  and lower half space ($z < 0$) of permittivity $\epsilon_{2}(\omega)$ and permeability $\mu_{2}(\omega)$. 
The two media are joined at interface $z=0$. SP pulses propagate at interface along x direction, and decay along z-direction. The two-level system (levels 1 and 2) interacting with SP pulses of frequency is placed above interface at position z.}
\end{figure}

These SP modes are confined to the interface plane and propagate in the x-direction along the interface with complex wave vector $\mathbf{\textit{K}}_{||}$. 
We shall assume that the propagation along 1D x-direction is facilitated by a channel or groove 
\cite{Zhang14} 
of width $L_y$ in the y-direction so as to direct the SP propagation along x-direction.
Shortly we will estimate the required width $L_y$ to channel the SP modes in the x-direction. 
The SP electric field amplitudes decay away in both sides with distance from the interface at $z = 0$. 
So the SP electric field $\mathbf{E}$ of a transverse mode of frequency $\omega$ satisfying the wave equation
$\nabla^2\mathbf{\textit{E}_m}+\omega^2\varepsilon_0\mu_0\
\varepsilon_m(\omega)\mu_m(\omega)\mathbf{\textit{E}_m}=0$
(m=1,2 for the two media), is of the form
$\mathbf{\textit{E}_1}=\mathbf{\textit{A}_1}e^{i(K_{||}x-\omega t)}e^{-k_1z}$
in the upper half space ($z > 0$), and 
$\mathbf{\textit{E}_2}=\mathbf{\textit{A}_2}e^{i(K_{||}x-\omega t)}e^{k_2z}$ 
in the lower space ($z < 0$).
Here the constants $\mathbf{\textit{A}_1}$ and $\mathbf{\textit{A}_2}$ can be determined from the boundary conditions.
Where $\varepsilon_0$ is the vacuum dielectric constant (or permittivity) and $\mu_0$ is the vacuum permeability, $c=1/\sqrt{\varepsilon_0\mu_0}$ is speed of light
in vacuum, $\varepsilon_m(\omega)$  is the dielectric function of the medium commonly designated in
the literature as electric permittivity, and $\mu_m(\omega)$ is medium magnetic permeability.
The wave numbers
$k_m=\sqrt{K_{||}^2-(\omega/c)^2\varepsilon_m(\omega)\mu_m(\omega)}$
 are the wave vector components
along z-direction normal to the interface characterized by positive real parts $Re[k_m]>0$  so that the SP field amplitudes decay away from interface. 
These SP modes are thus bound to interface and propagate at wave vector $\mathbf{K}_{||}$ parallel to interface.
Applications of appropriate boundary conditions at interface $z=0$, leads to the following condition for TM polarized SP modes:

\fla{
&k_1\varepsilon_2(\omega)+k_2\varepsilon_1(\omega)=0,
\nonumber
\\
&K_{||}=k_{||}+i\kappa=\frac{\omega}{c}\sqrt{\varepsilon_1\varepsilon_2\frac{\mu_1\varepsilon_2-\mu_2\varepsilon_1}{\varepsilon_2^2-\varepsilon_1^2}}.
}
\noindent
The case of TE modes can be analyzed along the same line with the exchange $\varepsilon\leftrightarrow\mu$. In this work we present analysis for the TM modes.

In these equations, the real part $k_{||}$ of the complex wave vector $K_{||}$ gives the dispersion relations for TM polarized SP modes, while the imaginary part $\kappa$ gives SP loss that determines the SP propagation distance $L_x=1/\kappa$ along the interface. 
The positive real parts of the wave numbers $k_m$, normal to interface give the skin or penetration depth of the fields into both media, which we take as our definition of field confinement and denote as $\zeta_m=1/Re[k_m]$. Smaller values of $\zeta_m=1/Re[k_m]$ indicate better confinement which means field can be confined to dielectric interface, and vice versa large $\zeta_m$ leads to poor confinement or may be de-confinement. The sp propagation distance $L_x$ and confinement are interrelated as we see shortly.   
Since real $k_{1,2}$ are positive, Eq. (1) is fulfilled when the electric permittivity of one of the two media has negative real part. 

To illustrate the basic dispersion and losses of these TM modes we take the example where the first medium is described by the pair ($\mu_1=1$  and $\varepsilon_1=2.5$) corresponding to a rare-earth crystal, while NIMM is modeled in the Drude model by the frequency dependent electric permittivity $\varepsilon_2(\omega)$ and magnetic permeability $\mu_2(\omega)$ as:

\fla{
\varepsilon_2(\omega)=1-\frac{\omega_e^2}{\omega(\omega+i\gamma_e)},
\mu_2(\omega)=1-\frac{\omega_h^2}{\omega(\omega+i\gamma_h)},
}
where $\omega_e$ is the electron plasma frequency usually in the ultraviolet region, $\gamma_e$ is the electric damping rate due to material losses, $\omega_h$ is the magnetic plasma frequency, $\gamma_h$ is the magnetic damping rate.

The dispersions are shown in Fig.2 where we display the mode frequency $\omega$ as a function of real part $k_{||}$ , and in Fig.3 we show losses as given by $\kappa(\omega)=Im[K_{||}(\omega)]$ of Eq. (1). 
The parameters are $\omega_e=1.37\times 10^{16}s^{-1}$,
$\gamma_e=2.73\times 10^{15}s^{-1}$ (for silver). 
Since the medium response to the magnetic component of the field is weaker than the electric component, we assume $\omega_h=\omega_e/2$  and $\gamma_h=\gamma_e/500$.
From Fig. 2, the dispersion relation is approximately linear in the frequency range below $\omega/\omega_e<0.35$ and losses are also highly reduced in this frequency range (Fig.3). 
Furthermore SP field is confined (as we see later) in this frequency range. So, for the rest of this work, we will assume the frequency range $0.05<\omega/\omega_e<0.35$, where dispersion is linear, losses are reduced and SP fields are confined as our operating working frequency range.

 \begin{figure}
\includegraphics[width=1.0\linewidth]{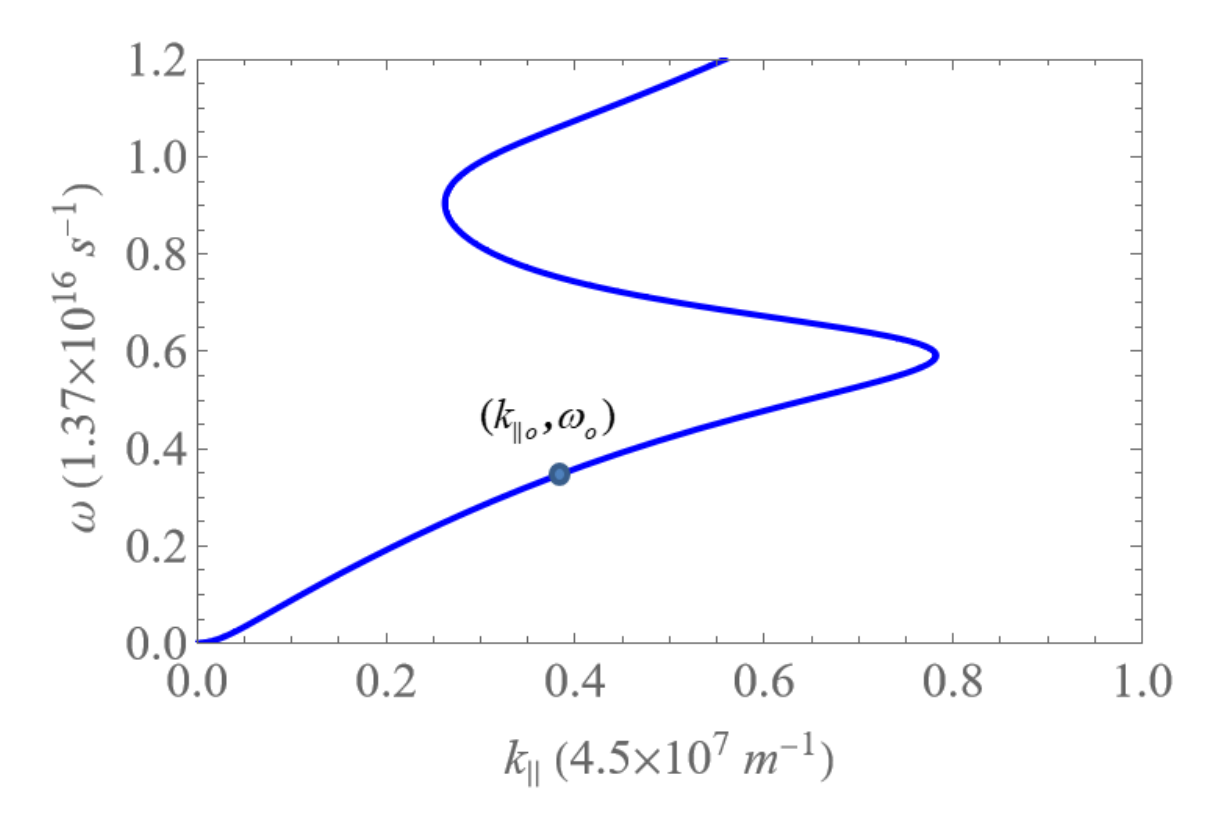}
\caption{Dispersion curves for TM modes. The point $\left(\text{k}_{||0},\omega_0\right)$ is shown. The linear segment in the lower branch of the curve is the part of interest to us.}
\end{figure}

\begin{figure}
\includegraphics[width=1.0\linewidth]{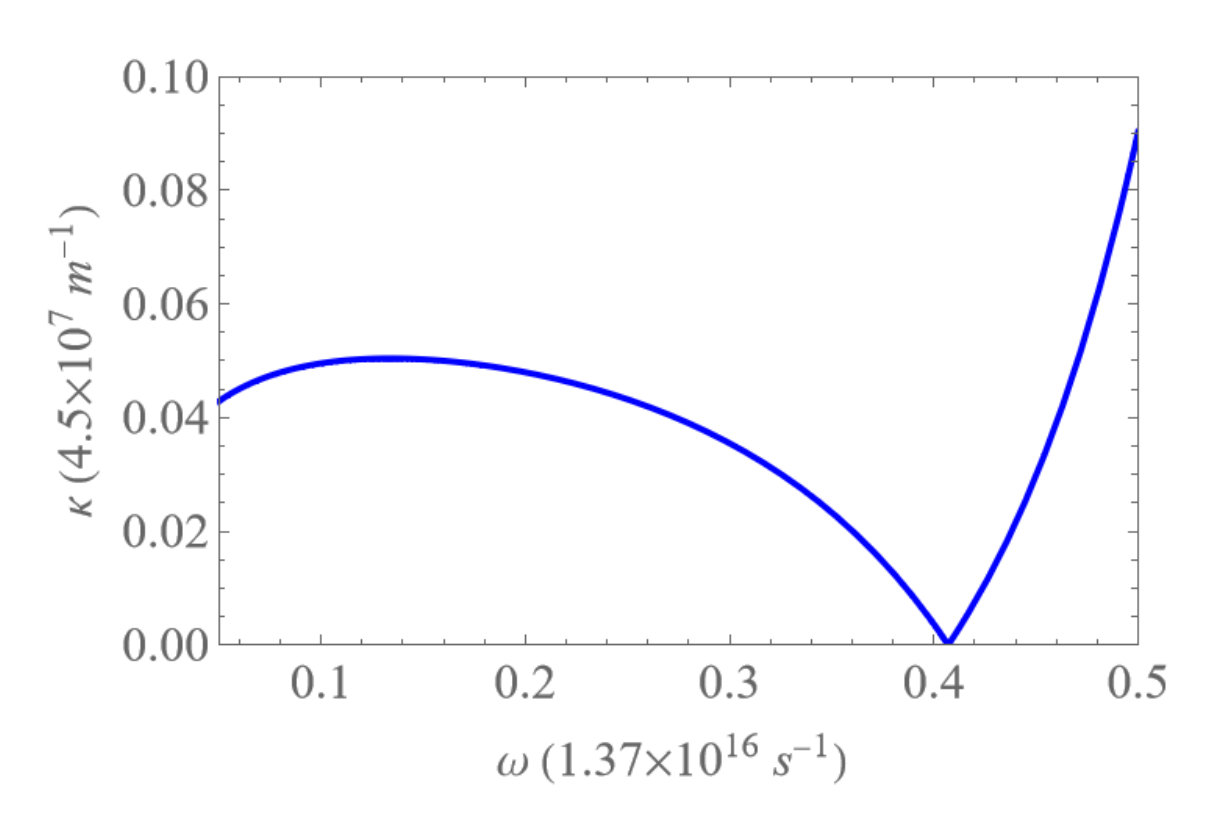}
\caption{ SP losses $\kappa(\omega)$ as given by the imaginary part of SP wave vector are shown as function of mode frequency $\omega$.}
\end{figure}

We now estimate the required channel width $L_y$ needed to direct SP propagation along x-direction by estimating the minimum uncertainty or deviation in the SP frequency ($\Delta\omega$) inside and outside the channel. 
The uncertainty in the SP energy is $\Delta E \Delta t \geq\hbar$  where $\Delta t\approx L_y/v_g$  is the time to traverse the channel width  $L_y$ at the surface plasmon group velocity $v_g$ calculated from the dispersion relation and shown in Fig.4 in dotted line in units of c.
 Then the SP energy deviation is $\Delta E \geq \hbar v_g/L_y$. Taking the channel width of order $400$ nm and, $v_g\approx0.8c$ (see Fig.4), the frequency deviation corresponding to this energy deviation is $\Delta\omega=0.5eV/\hbar$. 
For SP modes (without the channel) the frequency is usually of the order of the plasma frequency 
$\omega_e=1.37\times 10^{16}s^{-1}$   
(for silver for example), and thus the SP energy is
$ E=\hbar \omega = 9.2 eV$. 
Then for the channel width $L_y\approx 400$ nm, the frequency deviation is reduced by an amount (($0.5eV/\hbar)/(9.2 eV/\hbar)\approx 5\%$).
The frequency deviation is the difference between the SP frequency inside and outside the channel, so by taking the channel width of the order $400$ nm, this frequency deviation can be reduced and made as small as $5\%$ 
which means that the width $400 nm$ is large enough to neglect the frequency deviation and assume the SP frequency inside and outside the channel are practically the same and ignore the channel edge effects.

\section{Quantized SP fields }

We are concerned here with SP transport along \textbf{x} direction, so we shall need the quantized fields in order to quantify such effects. 
In the low loss range ($0.05<\omega/\omega_e<0.35$) $K_{||}=k_{||}$, and the SP quantization will determine the field amplitudes needed to couple to the two level atom that we need to consider in the next section. 
The quantization procedure leads to the following expressions for the single mode TM polarized SP field:

\fla{
\hat{\textbf{E}}_{k}(\textbf{r},t)=\textbf{E}_{0k}(z)\hat{a}_{k}(t)e^{ikx}+H.C.
}

\noindent where we use $k=k_{||}$ to simplify notations. 

 The TM mode field operators $\hat{a}_{k}(t)$, and  $\hat{a}^{\dagger}_{k}(t)$  of the plasmonic modes obey the usual equal time
 commutation relation $[\hat{a}_{k}(t),\hat{a}^{\dagger}_{k^{'}}(t)]=\delta_{k,k^{'}}$. 
The SP field amplitude $\textbf{E}_{0k}$   takes the form: 

\fla{
&{\textbf{E}_{0k}}(z)=N(\omega)\cdot
\nonumber
\\
&\left(\theta(z)(\hat{\textbf{x}}+i\hat{\textbf{z}}\frac{k}{k_1})e^{-k_1z}+
\theta(-z)(\hat{\textbf{x}}-i\hat{\textbf{z}}\frac{k}{k_1})e^{k_1z}\right).
}
where $\theta(z)$ is the Heaviside step function and

\fla{
\text{N}(\omega)= \sqrt{\frac{\hbar\omega}{2\pi\epsilon_0 L_y L_z(\omega)}},L_z(\omega)=D(\omega)+\frac{\omega^2}{c^2}S(\omega)
}

\fla{
\text{D}(\omega)=& \zeta_1\text{Re}\left(\frac{\partial(\omega\epsilon_1)}{\partial\omega}\right)\frac{|k_1|^2+|k|^2}{|k_1|^2}+
\nonumber
\\
& \zeta_2\text{Re}\left(\frac{\partial(\omega\epsilon_2)}{\partial\omega}\right)\frac{|k_2|^2+|k|^2}{|k_2|^2},
\\
\text{S}(\omega)=& \zeta_1\text{Re}\left(\frac{\partial(\omega\mu_1)}{\partial\omega}\right)|\frac{\epsilon_1}{k_1}|^2+\zeta_2\text{Re}\left(\frac{\partial(\omega\mu_2)}{\partial\omega}\right)|\frac{\epsilon_2}{k_2}|^2.
 }

In the above equations, the normalization factor $N(\omega)$ determines the field amplitude and is given in terms of various plasmonic parameters. The length $L_z$ is proportional to the confinement $\zeta_m=1/Re[k_m]$ and is determined by the physical properties of the NIMM medium such as by its permittivity's $\epsilon$  and permeability's $\mu$, given above
(see details in \cite{Kamli08,Moiseev10}). 

An important point to note in Eqs.(5-7) is that larger values of the wave numbers $Re[k_m]$ lead to smaller values for $\zeta_{m}=1/Re[k_m]$ and thus highly confined modes.
Likewise suppressed values of $Re[k_m]$ mean large values of $\zeta_{m}=1/Re[k_m]$ indicating poor confinement and thus large value of $L_z$. 
Fig.4 shows $\zeta_1=1/Re[k_1]$ (solid line) as a function of mode frequency. 
We can see from this figure and Fig.3 that there is a trade off between confinement and losses (or propagation distance $L_x$). 
In the frequency range where confinement is high,  we see that losses are also high leading to short SP propagation distances, while for poor confinement we see that losses are low and longer propagation distance. 
So we need to find the optimal confinement with optimal losses that suit practical needs. Appropriate choice of materials, i.e., adjusting the pairs  ($\epsilon_1$, $\mu_1$) and 
($\epsilon_2$, $\mu_2$), can lead to a decrease in $L_z$. 
The reduction of the interaction volume ($L_x L_y L_z$)due to confinement considerably enhances the field amplitude and Rabi frequency and thus leads to strong atom-field coupling. 

\section{SP pulse interaction with two level  atomic ensemble}

We consider the coupling of the SP pulse modes to an ensemble of two level atoms near a NIMM interface as shown in Fig. 1.
The total Hamiltonian of the SP pulse and two level atoms is  given as: 

\fla{\hat{H}=\hat{H}_a+\hat{H}_f+\hat{H}_{int},
}
where
\fla{\hat{H}_a=\frac{1}{2}\sum_j\hbar(\omega_0+\Delta_j)\sigma_z^j,
\nonumber
\hat{H}_f=\sum_k\hbar\omega_k \hat{a}_k^{\dagger}\hat{a}_k,
\\
\nonumber
\hat{H}_{int}=-\frac{1}{2}\sum_j\hbar\Omega(z_j)\sigma_{+}^j \hat{a}(x_j,t)e^{ikx_j}+H.C.,
}
where  $\hat{H}_a$  is the atomic ensemble Hamiltonian with central transition frequency $\omega_0$ and detuning $\Delta_j$  for the jth atom that are  inhomogeneously broadened by Gaussian function $G(\frac{\Delta}{\Delta_{in}})$ with spectral width $\Delta_{in}$,
$\hat{H}_f$  is the SP field part with mode frequency $\omega_k$, and $\hat{H}_{int}$  is the interaction Hamiltonian in the dipole approximation at the position of atom  $r_j=(x_j,z_j)$, and $\hbar$ is the reduced Planck constant. 
The two level atom operators are raising $\sigma_{+}^j(t)$, lowering $\sigma_{-}^j(t)$  operators and the inversion $\sigma_{z}^j(t)$. 
The SP field is given in Eqs (3)-(6). 
In this work we are interested in the transport of SP fields in space so it is more appropriate to work with SP field operator in the space formalism. 
We define the space representation SP field operators $\hat{a}(x,t)$ its Hermitian conjugate $\hat{a}^{\dagger}(x,t)$  in terms of their Fourier transforms operators $\hat{a}_k(t)$  and ($\hat{a}_k^{\dagger}(t)=(\hat{a}_k(t))^{\dagger}$ ) by the relations:

\fla{
\hat{a}(x,t)=&\frac{1}{\sqrt{2\pi}}\int_{-\infty}^{\infty}dk\hat{a}_k(t)e^{ikx}
\nonumber
\\
\hat{a}_k(t)=&\frac{1}{\sqrt{2\pi}}\int_{-\infty}^{\infty}dx\hat{a}(x,t)e^{-ikx}.
}
Furthermore for the operators  $\hat{a}(x,t)$  and $\sigma_{\pm}^j(t)$  we introduce the slowly varying operators $\tilde{a}(x,t)$, $\tilde{\sigma}_{\pm}^j(t)$ :

\fla{\hat{a}(x,t)=\tilde{a}(x,t)e^{-i(\omega_0 t-k_0 x)},
\nonumber
\\
\sigma_{\pm}^j(t)=\tilde{\sigma}_{\pm}^j(t)e^{\pm i(\omega_0 t-k_0 x)},
}
while $\sigma_{z}^j(t)=\tilde{\sigma}_{z}^j(t)$. Then the Heisenberg equation of motion 
$i\hbar \partial\hat{Q}(t)/\partial t=[\hat{Q}(t),\hat{H}] $
gives Maxwell-Bloch equations of motion for the slowly varying operators:

\fla{
&\left( \frac{\partial}{\partial t} + v_g \frac{\partial}{\partial x} \right) \tilde{a}(x,t) =
i\sqrt{\pi/2} \sum_{j} \Omega^{*}(z_j) \tilde{\sigma}_{-}^j(t)\delta(x-x_j),
\nonumber
\\
&\frac{\partial \tilde{\sigma}_{-}^j(t)}{\partial t} = -  i\Delta_{j}\tilde{\sigma}_{-}^j(t)-\frac{i}{2}\Omega(z_j)\sigma_{z}^j(t)\tilde{a}(x,t),
\nonumber
\\
&\frac{\partial \tilde{\sigma}_{z}^j(t)}{\partial t} = i\left[
\Omega(z_j)\sigma_{+}^j(t)\tilde{a}(x,t)
-\Omega^*(z_j)\sigma_{-}^j(t)\tilde{a}^{\dagger}(x,t)\right].
\label{eq::4}
}

where the complex Rabi frequency $\Omega(z_j)=2\textbf{d}_{21}^{\text{j}}\cdot \textbf{E}_{0}(z_j)/\hbar$, is a function of position z along z-direction normal to interface and mode frequency 
$\omega_k$, the stars indicates the complex conjugate ($\tilde{\sigma}_{+}k(t)=(\tilde{\sigma}_{-}(t))^{\dagger}$).
The SP field amplitude is given by the Eqs (3)-(6) and $\textbf{d}_{21}^{\text{j}}$ is a dipole moment of the atomic transition. 
The SP group velocity $v_g$=$\partial\omega/\partial\text{k}$, results from the expansion of  $\omega_k=\omega_0+(k-k_0)\partial\omega/\partial\text{k}$, (with $k=k_{||}$). 
In this expansion (see figure 2), $\omega_0=\omega_k|_{k=k_0}$ corresponding to the two level transition frequency center as in Fig.1, and $k_0$  is the corresponding wave number on the dispersion relation. 

To proceed with Eq.11 we shall assume that the SP coherent field is represented by a coherent state whose eigenvalues are complex numbers so that we can replace the SP field operator by a complex quantity of some amplitude and phase. Furthermore we assume the average of product of operators can be replaced by the product of averages i.e. 
$\langle \sigma_{z}^j(t)\tilde{a}^{\dagger}(x,t)\rangle\cong$
$\langle \sigma_{z}^j(t)\rangle\langle\tilde{a}^{\dagger}(x,t)\rangle=S_{z}^j(t)b(x,t)e^{-i\varphi(x)}$, 
$\langle \sigma_{z}^j(t)\tilde{a}(x,t)\rangle\cong$.
$S_{z}^j(t)b(x,t)e^{i\varphi(x)}$, 
$\langle \sigma_{\pm,z}^j(t)\rangle=S_{\pm,z}^j(t)$ where $b(x,t)$ and $\varphi(x)$ are the SP field amplitude (real value) and phase. 
Below we also consider only the symmetrical shape of the inhomogeneous broadening $G(\frac{\Delta}{\Delta_{in}})$. 
When calculating the macroscopic response in the field equation of (11), we replace summation with integration as follows:
\fla{
&\sum_{j} \Omega^{*}(z_j) 
S_{-}^j(t) \delta(x-x_j)=
\nonumber
\\
&\int dz \rho(x,z)\int_{-\infty}^{\infty}
d\Delta G(\frac{\Delta}{\Delta_{in}})  \Omega^{*}(z) S_{-}(\Delta, x,z,t).
}
\noindent
Integrating the first equation in (11) for SP pulse $\langle\tilde{a}(x,t)\rangle=b(x,t)e^{i\varphi(x)}$ in time $\int_{-\infty}^{\infty} dt...$ and assuming all atoms initially in their ground states, $S_{-}(\Delta, x,z,-\infty)=0$, so we get:

\fla{
&v_g \frac{\partial}{\partial x} \left(b(x,t)e^{i\varphi(x)}\right) =
i\sqrt{\pi/2}\int dz \rho(x,z)\cdot
\nonumber
\\
&\int_{-\infty}^{\infty}
d\Delta G(\frac{\Delta}{\Delta_{in}})  \Omega^{*}(z) \int_{-\infty}^{\infty} dt S_{-}(\Delta,x,z,t)=
\nonumber
\\
&i\sqrt{\pi/2} \int dz \rho(x.z)|\Omega(z)|^2\int_{-\infty}^{\infty}
d\Delta G(\frac{\Delta}{\Delta_{in}})  
\nonumber
\\
&(-i) \int_{-\infty}^{\infty} dt 
\frac{b(x,t)e^{i\varphi(x)}S_{z}(\Delta,x,z,t)}{\gamma+i\Delta}=\pi\sqrt{\pi/2} G(0)\cdot
\nonumber
\\
&\int dz \rho(x,z)|\Omega(z)|^2 \int_{-\infty}^{\infty} dt 
b(x,t)e^{i\varphi(x)}S_{z}(\Delta=0,x,z,t),
}
where, to describe the atomic response to the action of a SP pulse, we introduced a negligibly small decay constant of the atomic coherence $\gamma\rightarrow 0$, and defined the envelope area $\eta(x,t)=\int_{-\infty}^{t}b(x,t)dt$ and where $\eta(x)=\int_{-\infty}^{\infty}b(x,t)dt$,  appropriate for an ensemble of atoms interacting with  SP short pulse of TM polarization.

By using solution of Eq.(11) for $S_{z}(\Delta=0, x,z,t)=S_{z}^0 \cos \left(|\Omega(z)|\theta(x,t)\right)$ (where $S_{z}(0, x,z,-\infty)=S_{z}^0$) we obtain for Eq. (13):

\fla{
\frac{\partial}{\partial x} \left(\eta(x)e^{i\varphi(x)}\right)=\pi
\sqrt{\frac{\pi}{2}}\frac{S_{z}^0 G(0)}{v_g}e^{i\varphi(x)}
F(\eta(x))}

where
\fla{
F(\eta(x))=-\int dy dz \rho(x,z)|\Omega(z)|
\sin \left(|\Omega(z)|\eta(x)\right).
}

We see that $\varphi(x)=Const$ is a solution of (14).
The integration in Eq.(15) is evaluated over the cross section  (y-z plane) normal to the field propagation direction $\textbf{x}$. 
It is clear from this integral-differential equation that the pulse transport properties are determined from the form of atomic density $\rho(x,z)$ and z-dependence  of the Rabi frequency which has position and mode frequency dependence (suppressed so far).For TM polarized SP modes it is given from the SP field relations (4):

\fla{
\Omega(z,\omega)=\Omega_0(\omega) e^{-k_1 z} , \Omega_0(\omega)=2\frac{\text{N}(\omega)}{\hbar}\left(d_x+\text{i}d_z\frac{k}{k_1}\right).
}

for the dipole moment of atomic transition 
$\textbf{d}_{21}=\hat{\textbf{x}} d_x+\hat{\textbf{y}} d_y+\hat{\textbf{z}} d_z$, and the atom SP-field interaction takes place above the interface in medium 1. The Rabi frequency is composed of components parallel $d_x$ and normal $d_z$ to interface with modulus; $|\Omega(\omega_0)|=(2N(\omega)/\hbar) \sqrt{d_x^2+d_z^2|k/k_{1}|^2}$, and we have explicitly shown SP mode frequency dependence of the Rabi frequency.  

Now assuming the concentration of atoms to be constant $\rho(x,z)=\rho_0$, the 
function F in Eq(15) is given by the width of the channel $L_y$ times the integration over $z$, and the integral reduces to

\fla{
F(\eta(x))=-& L_y\rho_0 \int_{0}^{\infty} dz |\Omega_0(\omega)| e^{-k_1 z}\sin{\left(|\Omega_0(\omega)| e^{-k_1 z}\eta(x)\right)}
\nonumber
\\
=-&\frac{ L_y \rho_0 \sin^2 \left(|\Omega_0(\omega)|\eta(x)/2\right)}{\text{Re}[k_1] \eta(x)}.
 } 
 Thus Eqs.(14) and (17) result in
 
 \fla{
\frac{\partial}{\partial x} \eta(x)=-\alpha_0(\omega)
\frac{\sin^2 \left(|\Omega_0(\omega)|\eta(x)/2\right)}{\eta(x)},
}

\noindent
where $\alpha_0(\omega)=\pi\sqrt{\frac{\pi}{2}}\frac{\zeta_1 (\omega)L_y\rho_0G(0)}{v_g(\omega)}$ , with the closed analytical expression for   
the pulse area $\theta(x)=|\Omega_0(\omega)|\eta(x)$:
 \fla{
\text{x}=\text{x}_0-
\frac{1}{\alpha(\omega)}\left(T[\frac{\theta(x)}{2}] 
-T[\frac{\theta(x_0)}{2}]\right),
}

\noindent
where    $\alpha(\omega)=|\Omega_0(\omega)/2|^2\alpha_0(\omega)$,

 \fla{
T[y]=\text{ln}\sin[y]-y\cot[y].
}

\noindent
The "surface plasmonic area theorem" (SP area theorem) as given by Eqs.(18)  and its solution Eq.(19) are the main results here.
Eq. (18) has very rich spectrum of structure parameters, that can be utilized to explore its different facets; the modified pulse area  
$\theta(x)=|\Omega_0(\omega)|\eta(x)$  depends among many parameters on the SP modes confinement $\zeta_1(\omega)=1/Re[k_1]$, on the SP group velocity $v_g(\omega)$, the cross sectional area of propagation $L_y\zeta_1(\omega)$ , and on the modulus of Rabi frequency $|\Omega_0(\omega)|$. 

It is instructive at this point to estimate the magnitude of the Rabi frequency that quantifies the interaction of SP modes with resonant two level atoms. From Eq.5 we have $\Omega_{0}(\omega)=2d\sqrt{\omega/(2\pi\epsilon_{0}\hbar L_{y} L_{z})}$. For optical transition frequency we take 
$\omega=4\times10^{15}(s^{-1})$, and $L_z$ is of the order $\zeta_1$ which from Fig.4 is about 500nm. So with $L_y=400nm$ and dipole moment of order $d=2\times10^{-29}(C\cdot m)$, we get $\Omega_{0}(\omega)\approx4\times10^{8}(s^{-1})$. This quantity enters into the formula for the atomic absorption coefficient which we explore further. 
Fig.4 plots confinement and group velocity the two quantities that enter into the definition of absorption coefficient $\alpha$ as functions of mode frequency, and Fig.5 shows, together with ohmic losses $\kappa(\omega)$, the behavior of the atomic absorption coefficient $\alpha(\omega)$ taking into account the spectral dependence of the Rabi frequency $\Omega(\omega)$, the group velocity $v_g(\omega)$, and the spatial localization of surface plasmon $\zeta_1(\omega)$  near the interface.
It is seen that the resonance absorption coefficient can be more than three orders of magnitudes higher than ohmic plasmon losses  $\alpha(\omega)\gg \kappa(\omega)$ in 
a wide spectral range $0.05<\omega/\omega_e<0.35$, when using rare earth ions with a concentration $\rho=10^{14}m^{-2}$.
In this frequency range, the spatial localization of surface plasmons near the interface is about $\zeta_{1,max}\cong 500 nm$. 
For the validity of  using the SP area theorem, the resonant atoms must be located in a layer (see Fig.1) approximately no less than $\sim3\zeta_{1,max}$.
The possibility of increasing resonant absorption by using a higher atomic concentration requires special consideration in order to take into account the influence of interatomic interactions on the growth of the atomic phase relaxation in the hybrid system under consideration. It is worth noting that the pulse area $\theta(x)$ in the  SP area theorem is determined by the maximum of SP amplitude, which takes place near the interface surface. Accordingly, the enhanced interaction of SP field with resonant atoms is directly affected by the increase in the optical density of the resonant transition.

It is interesting to compare the properties of the SP area theorem and the  McCall-Hahn area theorem \cite{MacCall69,Ablowitz1974,Eberly98}  derived for the case of the interaction of a light pulse with two-level atoms in a free space,  which has the general solution:

\fla{\text{x}=\text{x}_0-\frac{1}{\alpha}\text{ln}\left(\frac{\text{tan}[|\Omega_0|\theta(x)/2]}{\text{tan}[|\Omega_0|\theta(x_0)/2]}\right).
}
\noindent                                
In Figs.6,7 we compare the two area theorems as given by Eq.(19) and Eq.(21) for the same initial pulse areas. 

As can be seen in both cases, the evolution of the pulse areas of the input signals leads to the formation of $2n\pi$ pulses propagating over long distances in an optically dense medium.
At the same time, there are significant differences in the dynamics of the formation of these $2\pi$ pulses in these two cases. 
In contrast to the solution of McCall-Hahn area theorem \cite{MacCall69}, SP-theorem  does not have the bifurcation points $\theta (x_0)=(2n+1)\pi$. 
In addition, the $2 \pi$  SP pulses arising in the theorem are unstable: even with a slight decrease in the pulse area relative to $2n \pi$ ($\theta (x_0)=2n \pi-\epsilon$, $\epsilon\ll 1$), its subsequent evolution leads to the state $\theta(x\rightarrow \infty)\rightarrow 2(n-1)\pi$, as it is seen in Fig.7.
Moreover,
unlike the formation of $2\pi$ pulses (solitons) in free space, an increase in the total pulse area of the input SP pulse $\theta(0)>2n\pi$, ($n>2,3,...$) lengthens the formation time of independent $2 \pi$ pulses.
SP area theorem does not provide information about the temporal shape of $2 \pi$ SP pulses and the dynamics of their splitting in the medium, analysis of this issue is beyond the scope of this work.

\begin{figure}
\includegraphics[width=1.0\linewidth]{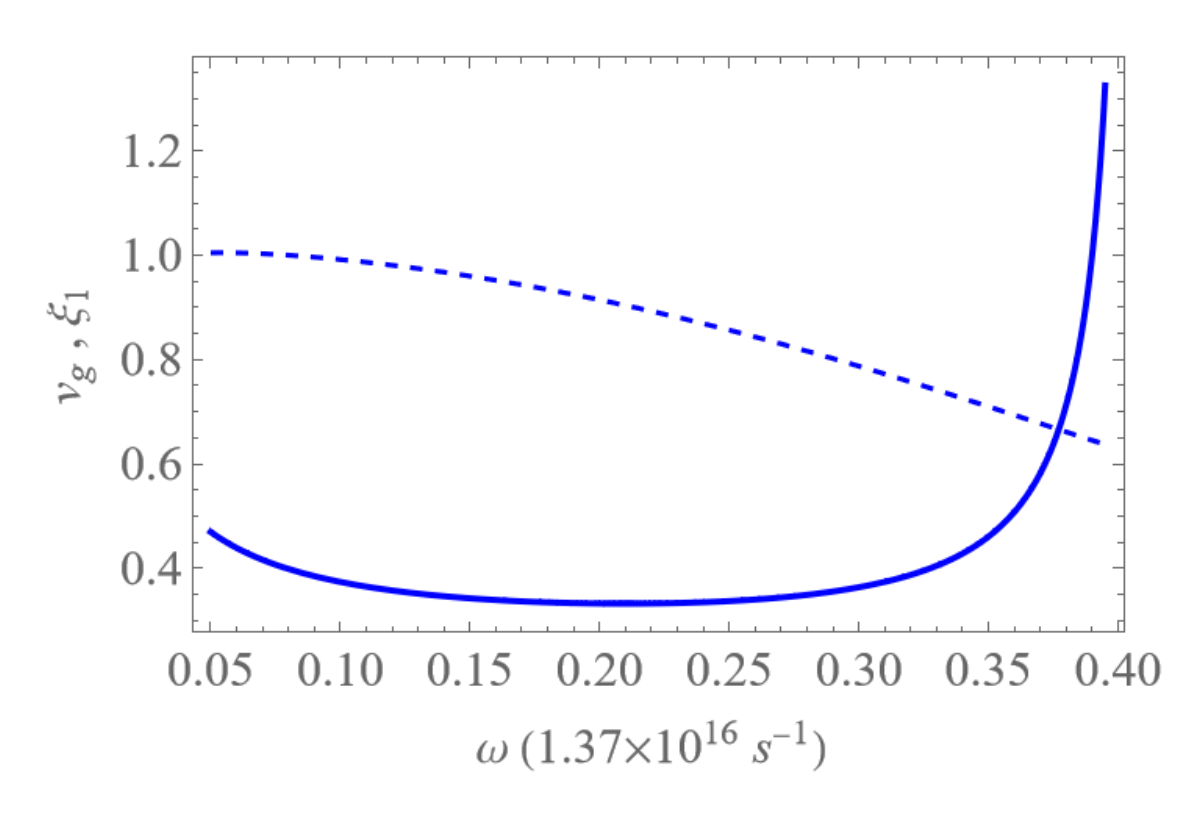}
\caption{The solid line is confinement $\zeta_1(\omega)=1/Re[k_1]$ $(\mu m)$, and dotted line is SP group velocity in units of c.}  
\end{figure}

\begin{figure}
\includegraphics[width=1.0\linewidth]{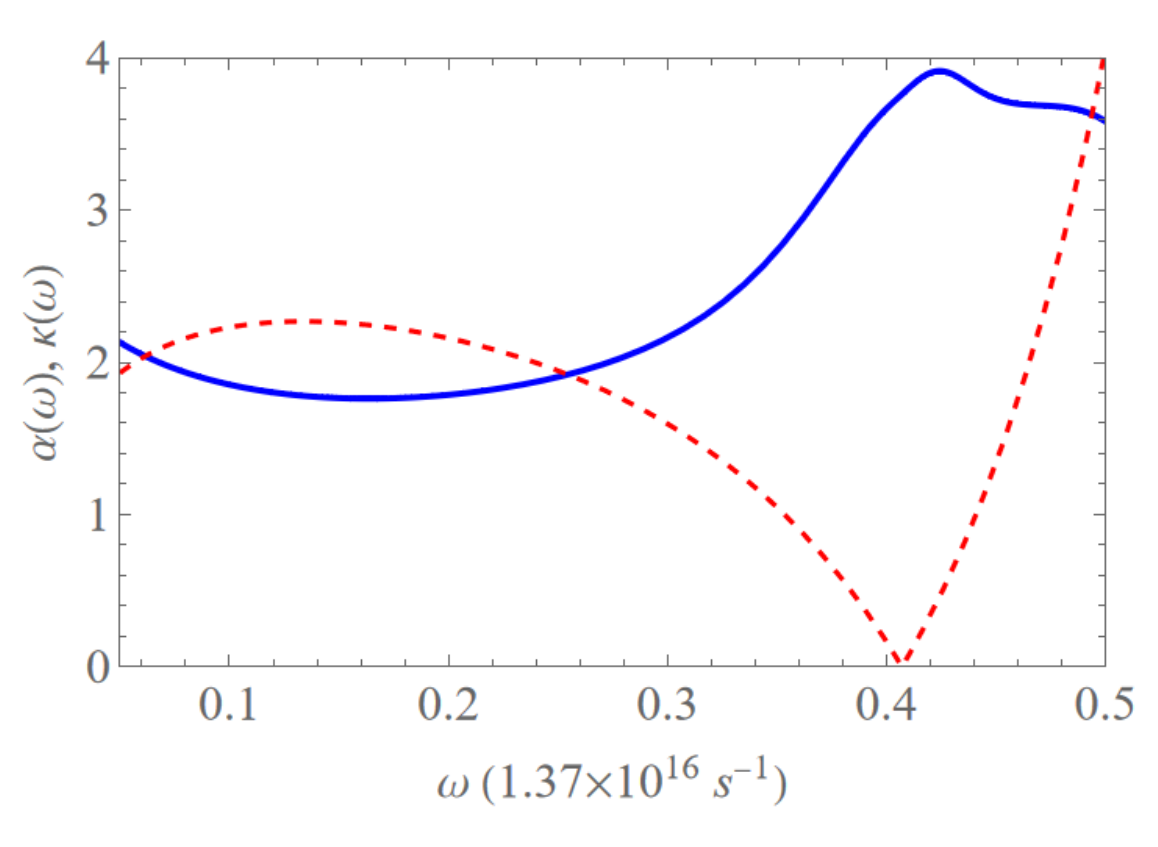}
\caption{Atomic absorption coefficient $\alpha(\omega)$ (solid line in units of $10^9/m$) and SP losses $\kappa(\omega)$ (dotted line in units of $10^6/m$) as functions of $\omega$.} 
\end{figure}

\begin{figure}
\includegraphics[width=1.0\linewidth]{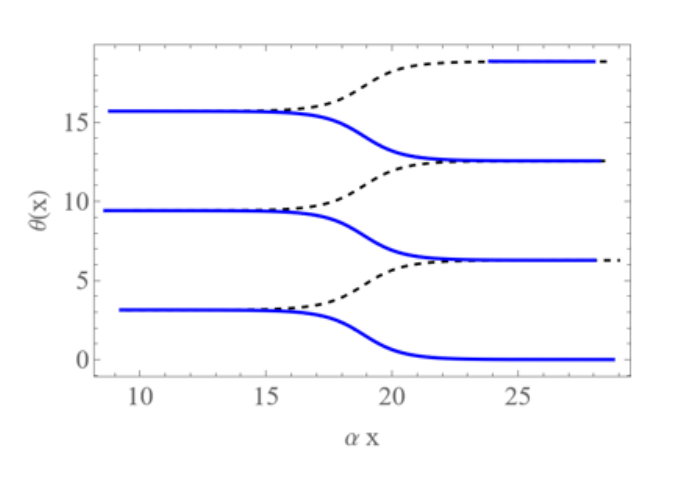}
\caption{MacCall-Hahn area theorem Eq.21 \cite{MacCall69} is shown for initial conditions $\theta(x_0)=0.9\pi, 2.9 \pi, 4.9 \pi$ (solid line) and $\theta(x_0)=1.1\pi, 3.1 \pi, 5.1 \pi$ (dotted line).} 
\end{figure}

\begin{figure}
\includegraphics[width=1.0\linewidth]{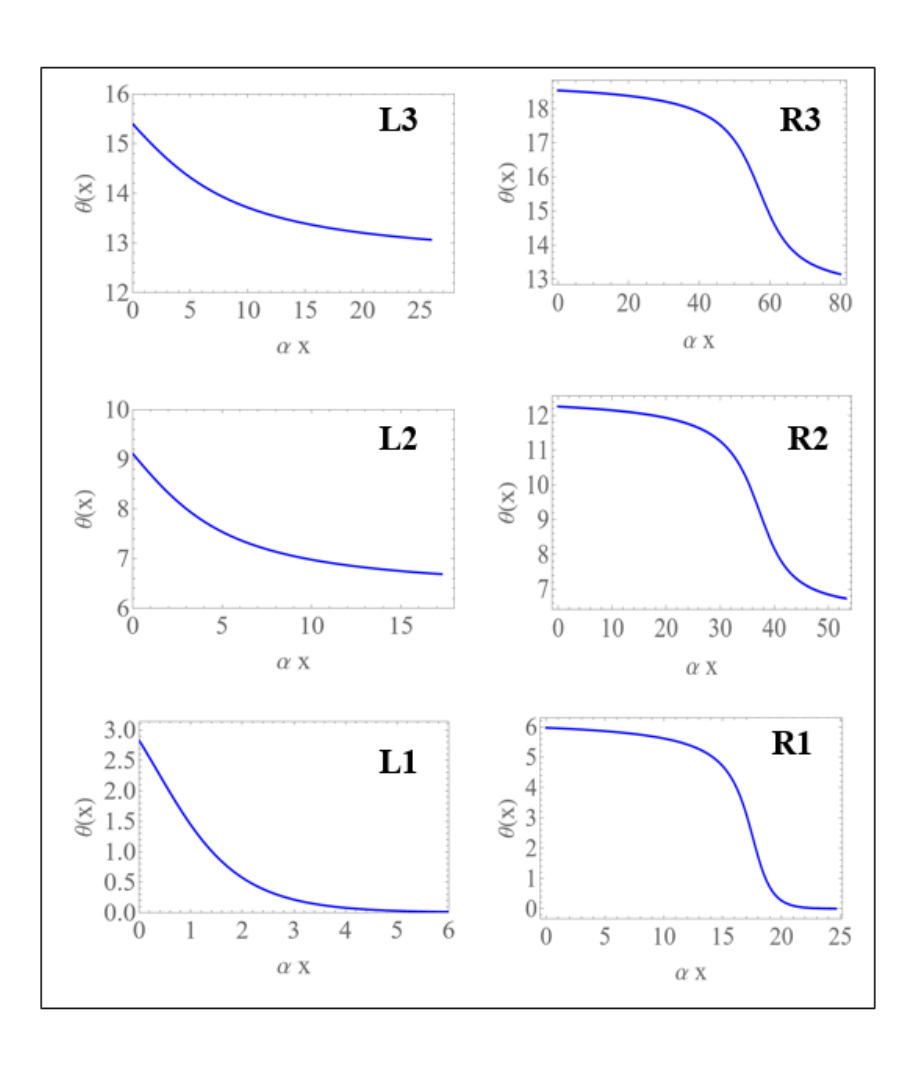}

\caption{The SP area theorem (Eq.19) for the initial conditions: (L1) $\theta(x_0)=0.9\pi$, (L2) $2.9\pi$, (L3) $4.9\pi$, (R1) $1.98\pi$, (R2) $3.98\pi$, and (R3) $5.98\pi$.} 
\end{figure} 

\section{Discussion and conclusion} 

The application of the MacCall-Hahn area theorem made it possible to study the general nonlinear patterns of resonant interactions between the light pulses and coherent resonant atomic ensembles in free space \cite{MacCall69,Ablowitz1974,Eberly98}, waveguide \cite{Moiseev23} and resonators \cite{Thierry2014,Moiseev22}. 
In this work, we developed the  pulse area approach for interaction of surface plasmon fields with two-level atoms. 
The derived SP area theorem Eq. (18) with its closed analytical solution (19) demonstrated a formation of long propagating $2\pi$ SP-pulses.
The long time in the formation of $2\pi$ SP-pulses with an increase in their number  
is a characteristic property of the SP area theorem, which becomes understandable given the influence of many parameters in the nonlinear interaction (resonance absorption coefficient) for atoms located at different distances from the interface.
At the same time, the very possibility of the appearance of $2\pi$ SP-pulses under conditions of a strong change in the amplitude of the SP-field for different atoms (with different distances z from the interface) remains not fully understood and even surprising for the strong nonlinear resonant interaction of SP fields with atoms.
Thus, it seems unlikely that the $2\pi$ SP-pulse propagates in an optically dense medium while maintaining its temporal shape.
Elucidation of the spatio-temporal structure of such $2 \pi$ SP-pulses and the features of their interaction with a resonant atomic ensemble will make it possible to understand the possibility of their occurrence and the ability to propagate over long distances. 
We note that despite significant differences in physical properties, Eq. (19) of the  SP area theorem  resembles the pulse area theorem in a single-mode optical waveguide \cite{Moiseev23}. 
The qualitative behaviours of the two cases may be traced back to the fact that both pulses have decaying field amplitudes that enter into the Rabi frequency and pulse areas. 
However for the considered single waveguide pulses  \cite{Moiseev23} the decay has the form of a Gaussian or Bessel function of the first kind whereas for SP-pulses decays are exponential and the nature of these decays demonstrates the significance differences between the two cases. However, this resemblance in the theoretical description  makes it possible to uniformly describe various effects of nonstationary nonlinear coherent interaction of structured light pulses  and SP fields with resonant atomic ensembles. 
Herein, due to the strong interaction of the SP-mode with resonant atoms, the formation of $2\pi$ SP-pulses will be detected experimentally at an  atomic concentration lower than in the case of the formation of optical solitons in a resonant medium
\cite{Allen75,MacCall69,Lamb71,Eberly98}.
It is worth noting that, unlike the very different behavior of the electromagnetic field of light modes in the cross section of various optical waveguides, the electromagnetic field of SP-fields usually decreases exponentially with distance from the interface.
This circumstance makes the SP area theorem important for describing the resonant interaction of atoms with SP-fields not only on NIMMs, but also on the noble metals and other materials. 
Graphene may be particularly interesting due to the very low attenuation of SP-fields and the large compactness of the devices being created.
We believe the $2\pi$  SP-pulses with their interesting properties outlined above are useful for further theoretical studies and experimental investigations.
We anticipate the SP area theorem can be helpful for studies of resonant interaction of SP-pulse with resonant two- and multi-level atoms, and various effects such as Dicke superradiance of SP-fields 
\cite{Pustovit09,Oulton09}, 
SP-lasing \cite{Bordo21,Seo23},
description  of SP-echo effect \cite{Mois2010} and SP-echo spectroscopy of atoms on the surface, microscopic transport of information via SP-fields, all of them deserve independent research. 

\section{ACKNOWLEDGMENTS}

 Sergey.A.M. was supported by Ministry of Education and Science of the Russian Federation (Reg. number NIOKRT 121020400113-1).

\bibliography{main}

\end{document}